\documentclass[sigconf,nonacm]{acmart}
\usepackage{enumitem}

\usepackage{graphicx}
\usepackage{soul}
\usepackage[dvipsnames]{xcolor}
\usepackage{multirow}

\usepackage[most]{tcolorbox}
\tcbset{
  colback=gray!5,
  colframe=gray!60,
  boxrule=0.5pt,
  arc=2pt,
  left=6pt,
  right=6pt,
  top=6pt,
  bottom=6pt
}
\usepackage{booktabs} 
\usepackage{listings}

\AtBeginDocument{%
  }

\setcopyright{none}
\begin{document}

%% COMMENTS
\definecolor{lightgray}{gray}{0.9}
\definecolor{lightblue}{rgb}{0.9,0.9,1}
\definecolor{LightMagenta}{rgb}{1,0.5,1}
\definecolor{red}{rgb}{1,0,0}
\definecolor{brightgreen}{rgb}{0.4, 1.0, 0.0}

\newcommand\couldremove[1]{{\color{lightgray} #1}}
\newcommand{\remove}[1]{}
\newcommand{\move}[2]{ {\textcolor{Purple}{ \bf --- MOVE #1: --- }} {\textcolor{Orchid}{#2}} }

%% FOR COMMENTS & EDITS
\newcommand{\hlc}[2][yellow]{ {\sethlcolor{#1} \hl{#2}} }
\newcommand\note[1]{\hlc[SkyBlue]{-- #1 --}} % highlighted notes of other colors.
% For colors info from xcolor package, check out:
% http://en.wikibooks.org/wiki/LaTeX/Colors

\newcommand\mynote[1]{\hlc[yellow]{#1}}
\newcommand\tingjun[1]{\hlc[yellow]{TC: #1}}
\newcommand\zhihui[1]{\hlc[LightMagenta]{ZG: #1}}
\newcommand\tom[1]{\hlc[brightgreen]{TOM: #1}}
\newcommand\andy[1]{\hlc[pink]{Andy: #1}}
\newcommand\bill[1]{\hlc[cyan!50]{Bill: #1}}
\newcommand\change[1]{{\color{blue} {#1}}}

% Revision highlights
\newcommand{\TODO}[1]{\textcolor{red}{#1}}
\newcommand{\revise}[1]{\textcolor{blue}{#1}}

% \newcommand\mynote[1]{\hlc[yellow]{#1}}
% \newcommand\tingjun[1]{\hlc[yellow]{TC: #1}}
% \newcommand\zhihui[1]{\hlc[LightMagenta]{ZG: #1}}
% \newcommand\tom[1]{\hlc[brightgreen]{TOM: #1}}
% % \newcommand\xiao[1]{\hlc[yellow]{XZ: #1}}
% \newcommand\change[1]{{\color{blue} {#1}}}

\newcommand{\myparatight}[1]{\vspace{0.5ex}\noindent\textbf{#1~~}}

\newcommand{\cmark}{\ding{51}}%
\newcommand{\xmark}{\ding{55}}%
\newcommand{\greencheck}{\color[HTML]{3C8031}{\cmark}}
\newcommand{\redcross}{\color[HTML]{ED1B23}{\xmark}}
\newcommand{\greenno}{\color[HTML]{3C8031}{\textbf{No}}}
\newcommand{\redyes}{\color[HTML]{ED1B23}{\textbf{Yes}}}
\newcommand{\greenlow}{\color[HTML]{3C8031}{\textbf{Low}}}
\newcommand{\redhigh}{\color[HTML]{ED1B23}{\textbf{High}}}

\newcommand*\circledwhite[1]{\tikz[baseline=(char.base)]{
            \node[shape=circle,draw,inner sep=0.6pt] (char) {\scriptsize{#1}};}}

\newcommand*\circled[1]{\tikz[baseline=(char.base)]{
            \node[shape=circle,draw,fill=black,text=white,inner sep=1pt] (char) {\scriptsize{#1}};}}

%% paper-specific variable/notation
\newcommand{\name}{IoT-SkillsBench}
\newcommand{\namebf}{\textbf{IoT-SkillsBench}}

%% code
\newcommand{\myCodeShort}[1]{\texttt{\small{#1}}}

\newcommand{\agora}{Agora}
\newcommand{\agorabf}{\textbf{Agora}}

\newcommand{\armavec}{\sf Savannah-mc (arma-vec)}
\newcommand{\armacube}{\sf Savannah-mc (arma-cube)}

%%This is a special cell that allows using \\ inside the cell to add a new line
\newcommand{\specialcell}[2][c]{%
  \begin{tabular}[#1]{@{}c@{}}#2\end{tabular}}

\newcommand{\iu}{{j}}

\newcommand{\littlesum}{\mathop{\textstyle\sum}}
\newcommand{\littleint}{\mathop{\textstyle\int}}

% config
\newcommand{\siso}{SISO}
\newcommand{\mimoTwoByTwo}{2$\times$2}
\newcommand{\mimoFourByFour}{4$\times$4}

% lib
\newcommand{\bbdev}{\textsf{bbdev}}

% PHY
\newcommand{\scs}{{\Delta f}}
\newcommand{\scNum}{N_{\textrm{sc}}}
\newcommand{\sampRate}{F_{\textrm{s}}}
\newcommand{\fftSize}{N_{\textrm{FFT}}}
\newcommand{\chMat}{{\textbf{H}}}
\newcommand{\chVec}{{\textbf{h}}}
\newcommand{\precodeMat}{{\textbf{P}}}

% units
\newcommand{\usec}{$\mu$s} % us or usec
\newcommand{\msec}{ms}     % ms or msec

% DSP stages
\newcommand{\fft}{\textsf{fft}}
\newcommand{\ifft}{\textsf{ifft}}
\newcommand{\csi}{\textsf{csi}}
\newcommand{\precode}{\textsf{precode}}
\newcommand{\encode}{\textsf{enc}}
\newcommand{\decode}{\textsf{dec}}
\newcommand{\modul}{\textsf{modul}}
\newcommand{\demod}{\textsf{demod}}
\newcommand{\equal}{\textsf{equal}}

% LDPC parameters
\newcommand{\tbSize}{T}
\newcommand{\tbCrcSize}{T_{\textrm{crc}}}
\newcommand{\cbSize}{K_{\textrm{cb}}}
\newcommand{\cbNum}{N_{\textrm{cb}}}
\newcommand{\liftingSize}{Z_{c}}
\newcommand{\liftingSizeSet}{\mathbf{\Theta}}
\newcommand{\fillerBitNum}{N_{\textrm{filler}}}

\newcommand{\codeRate}{R}

\newcommand{\throughput}{Tp}
\newcommand{\codingTime}{t}
\newcommand{\informationBits}{K'}

% Thresholds
\newcommand{\thresPower}{\theta_{\textrm{Power}}}
\newcommand{\thresOtsu}{\theta_{\textrm{Otsu}}}
\newcommand{\thresPSD}{\theta_{\textrm{PSD}}}
\newcommand{\thresIoU}{\theta_{\textrm{IoU}}}

% Real-time parameters
\newcommand{\tput}{\lambda} % throughput
\newcommand{\lat}{l} % latency

% MO parameters
\newcommand{\se}{E}

% OTFS parameters
\newcommand{\numDelayBin}{M}
\newcommand{\numDopplerBin}{N}
\newcommand{\delay}{\tau}
\newcommand{\DopplerShift}{\nu}
\newcommand{\delayRes}{\Delta\delay} % delay resolution
\newcommand{\DopplerRes}{\Delta\DopplerShift} % Doppler resolution
\newcommand{\pathIdx}{p}
\newcommand{\pathDelay}{\tau_{\pathIdx}}
\newcommand{\pathDoppler}{\nu_{\pathIdx}}
\newcommand{\delayIdx}{k}
\newcommand{\DopplerIdx}{l}
\newcommand{\pilotDelayIdx}{K_0}
\newcommand{\pilotDopplerIdx}{L_0}
\newcommand{\pathDelayIdx}{{\delayIdx}_{\pathIdx}}
\newcommand{\pathDopplerIdx}{{\DopplerIdx}_{\pathIdx}}

\newcommand{\chEleDD}{H_{\textrm{dd}}}
\newcommand{\chMatDD}{\textbf{H}_{\textrm{dd}}}
\newcommand{\chEleDDEst}{\widehat{H}_{\textrm{dd}}}
\newcommand{\chMatDDEst}{\widehat{\textbf{H}}_{\textrm{dd}}}
\newcommand{\chMatDDEstHerm}{\widehat{\textbf{H}}^\textrm{H}_{\textrm{dd}}}

% OTFS data structure parameters
\newcommand{\chMatDDRowIdx}{i}
\newcommand{\chMatDDColIdx}{j}
\newcommand{\chMatDDColIdxMapped}[1]{{\chMatDDColIdx}_{\pathIdx}(#1)}
\newcommand{\chMatDDColIdxMappedRowIdx}{\chMatDDColIdxMapped{\chMatDDRowIdx}}
\newcommand{\chMatDDRowIdxMapped}[1]{{\chMatDDRowIdx}_{\pathIdx}(#1)}
\newcommand{\chMatDDRowIdxMappedRowIdx}{\chMatDDRowIdxMapped{\chMatDDColIdx}}
\newcommand{\sparseCoe}{D}
\newcommand{\mvmCoe}[2]{\sparseCoe_{{#1},{#2}}}
\newcommand{\coePathRowIdx}{\mvmCoe{\pathIdx}{\chMatDDRowIdx}}
\newcommand{\chMatDDColDelayBinOffset}{{d}_{\delayIdx}(\pathIdx)}
\newcommand{\chMatDDColDopplerBinOffset}{{d}_{\DopplerIdx}(\pathIdx)}
\newcommand{\chPathGain}{h_{\pathIdx}}

\newcommand{\chNoiseDDVec}{\textbf{w}_{\textrm{dd}}}
\newcommand{\chNoiseDDMat}{\textbf{W}_{\textrm{dd}}}
\newcommand{\chNoiseDDMatEle}{W_{\textrm{dd}}}
\newcommand{\chNoiseTD}{w}

\newcommand{\chEleMMSE}{H_{\textrm{LMMSE}}}
\newcommand{\chMatMMSE}{\textbf{H}_{\textrm{LMMSE}}}

\newcommand{\chResponse}{h_{\textrm{eff}}}

\newcommand{\chEffEstMat}{\hat{\mathbf{h}}_{\textrm{eff}}}
\newcommand{\chEffEstMatEle}{\hat{h}_{\textrm{eff}}}
\newcommand{\chEffMat}{\mathbf{h}_{\textrm{eff}}}
\newcommand{\chEffMatEle}{h_{\textrm{eff}}}

\newcommand{\sigTxDD}{X_{\textrm{dd}}}
\newcommand{\sigTxDDVec}{\mathbf{x}_{\textrm{dd}}}
\newcommand{\sigTxDDVecEle}{x_{\textrm{dd}}}
\newcommand{\sigTxDDVecEst}{\widehat{\mathbf{x}}_{\textrm{dd}}}
\newcommand{\sigTxDDVecEstDemodEle}{\breve{x}_{\textrm{dd}}}
\newcommand{\sigTxDDMatEle}{X_{\textrm{dd}}}
\newcommand{\sigTxDDMat}{\mathbf{X}_{\textrm{dd}}}
\newcommand{\sigTxDDMatEst}{\widehat{\mathbf{X}}_{\textrm{dd}}}

\newcommand{\sigTxTdVecEle}{x}
\newcommand{\sigTxTdVec}{\mathbf{x}}
\newcommand{\sigTxTdMatEle}{X}
\newcommand{\sigTxTdMat}{\mathbf{X}}

\newcommand{\sigRxDD}{Y_{\textrm{dd}}}
\newcommand{\sigRxDDVec}{\mathbf{y}_{\textrm{dd}}}
\newcommand{\sigRxDDMatEle}{Y_{\textrm{dd}}}
\newcommand{\sigRxDDMat}{\mathbf{Y}_{\textrm{dd}}}
\newcommand{\sigRxTd}{y}
\newcommand{\sigRxTdVec}{\mathbf{y}}
\newcommand{\sigRxTdMat}{\mathbf{Y}}

\newcommand{\numax}{\nu_{\text{max}}}

\newcommand{\bandwidth}{B}
\newcommand{\frameTime}{T}
\newcommand{\numPath}{P} % corresponds to the number of non-zero entries in the channel matrix

\newcommand{\heffThres}{\theta}
\newcommand{\chPathThresSet}{\widehat{\Omega}}

% OTFS CGA parameters
\newcommand{\noiseCovMat}{\mathbf{R}_{n}}
\newcommand{\cgaIters}{K}
\newcommand{\sqResNorm}{c_\textrm{norm}}

% OTFS exponential terms
\newcommand{\kernelFftMatEle}{E_{\text{Zak}}}
\newcommand{\kernelFftMat}{\mathbf{E}_{\text{Zak}}}
\newcommand{\twistMatEle}{E_{\text{twist}}}
\newcommand{\twistMat}{\mathbf{E}_{\text{twist}}}

% OTFS indexing
\newcommand{\delayIdxSet}{\mathcal{\numDelayBin}}
\newcommand{\DopplerIdxSet}{\mathcal{\numDopplerBin}}
\newcommand{\pathIdxSet}{\mathcal{\numPath}}
\newcommand{\gridFlatIdxSet}{\mathcal{G}} % index set for flattened grid (0, 1, ..., mn-1)

% Other helpful
\newcommand{\jimg}{\mathrm{j}}
\newcommand{\defn}{\triangleq}
\newcommand{\mymod}[2]{\langle #1 \rangle_{#2}}

\newcommand{\myAbs}[1]{\left|{#1}\right|}
\newcommand{\myAng}[1]{\angle{#1}}
\newcommand{\myConjugate}[1]{{#1}^{*}}
\newcommand{\myTranspose}[1]{{#1}^{\top}}
\newcommand{\myHermitian}[1]{{#1}^{H}}
\newcommand{\myIsFunc}[1]{\mathbf{1}\{#1\}}

% Angle
\newcommand{\AoD}{\phi}
\newcommand{\AoDVec}{\bm{\upphi}}
\newcommand{\AoDbf}{\boldsymbol\phi}
\newcommand{\AoDDirectional}{\Phi}
\newcommand{\az}{\phi}
\newcommand{\azVec}{\bm{\upphi}}
\newcommand{\azVecUE}{\bm{\upphi}_{\textrm{UE}}}
\newcommand{\azbf}{\boldsymbol\phi}
\newcommand{\el}{\psi}
\newcommand{\elbf}{\boldsymbol\psi}

% Element Setup
\newcommand{\ElemComp}{w}
\newcommand{\ElemCompbf}{\mathbf{w}}
\newcommand{\ElemCompNew}{w^\prime}
\newcommand{\ElemCompNewbf}{\mathbf{w}^\prime}
\newcommand{\ElemAmp}{A}
\newcommand{\ElemAmpbf}{\mathbf{A}}
\newcommand{\ElemPhase}{\theta}
\newcommand{\ElemPhasebf}{\boldsymbol\theta}
\newcommand{\steer}{s}
\newcommand{\steerVec}{\mathbf{s}}
\newcommand{\steermat}{\mathbf{S}}
\newcommand{\beamPattern}{BP}

\newcommand{\bw}{B}
\newcommand{\carrierFreq}{f_{c}}
\newcommand{\carrierWave}{\lambda}

\newcommand{\csiMat}{\mathbf{H}}

\newcommand{\ASA}[2]{\textrm{ASA}({#1},{#2})}
\newcommand{\antNum}{N}
\newcommand{\antIdx}{n}
\newcommand{\antDist}{d}
\newcommand{\subarrayNum}{M}
\newcommand{\subarraySet}{\mathcal{M}}
\newcommand{\subarrayIdx}{m}
\newcommand{\subarrayAntNum}{N_{s}}
\newcommand{\subarrayAntIdx}{n}
\newcommand{\subarrayAntDist}{d}

\newcommand{\setSubarray}{\mathcal{A}}
\newcommand{\subarrayAlloc}{a}
\newcommand{\subarrayAllocVec}{\mathbf{a}}
\newcommand{\subarrayAllocMat}{\mathbf{A}}
\newcommand{\subarrayAllocSet}{\mathbb{A}}

% beamforming
\newcommand{\bfWeight}{w}
\newcommand{\bfWeightVec}{\mathbf{w}}
\newcommand{\bfAmp}{A}
\newcommand{\bfAmpVec}{\mathbf{A}}
\newcommand{\bfPhase}{\theta}
\newcommand{\bfPhaseVec}{\boldsymbol{\theta}}
\newcommand{\bfGain}{g}
\newcommand{\bfGainSig}[1]{g^{\textrm{sig}}_{#1}}
\newcommand{\bfGainInt}[2]{g^{\textrm{int}}_{{#1}\rightarrow{#2}}}

\newcommand{\power}{\mathcal{P}}
\newcommand{\powerSignal}{\mathcal{S}}
\newcommand{\powerSignalDiff}{d\mathcal{S}}
\newcommand{\powerInterf}{\mathcal{I}}
\newcommand{\powerInterfDiff}{d\mathcal{I}}
\newcommand{\powerNoise}{N}

% network model
\newcommand{\userNum}{U}
\newcommand{\userIdx}{u}
\newcommand{\userSet}{\mathcal{U}}

\newcommand{\userNumSub}{K}

\newcommand{\userSelected}{k}
\newcommand{\userSelectedNum}{K}
\newcommand{\userSelectedSet}{\mathcal{K}}

\newcommand{\userAngle}{\phi}
\newcommand{\userWeight}{\alpha}

\newcommand{\baseSNR}{\gamma}
\newcommand{\SNR}{\mathsf{SNR}}
\newcommand{\SNRMax}{\mathsf{SNR}^{\textrm{max}}}
\newcommand{\SINR}{\mathsf{SINR}}
\newcommand{\SINRMax}{\mathsf{SINR}^{\textrm{max}}}
\newcommand{\Capacity}{T}
\newcommand{\Rate}{R}
\newcommand{\RateMax}{\Rate^{\textrm{max}}}
\newcommand{\RateAvg}{\widebar{\Rate}}
\newcommand{\CapacityMax}{\Tilde{T}}
\newcommand{\suppress}{\alpha}

% Proportional Fair
\newcommand{\RateHist}{\widebar{\Rate}}

\newcommand{\past}{p}
\newcommand{\decay}{\beta}

\newcommand{\RateMean}{\Bar{R}}
\newcommand{\JFI}{\mathsf{JFI}}

\newcommand{\bigO}{\mathcal{O}} % Open at top left
\newcommand{\conv}{\ast}
% \newcommand{\conv}{\circledast}
% \newcommand{\conv}{\star}


% \title{IoT-SkillsBench: Skill-Grounded Agents for Embedded System Design}
% \title{Practical IoT Systems Development Using Skilled AI Agents}
\title{Skilled AI Agents for Embedded and IoT Systems Development}

\author{Yiming Li, Yuhan Cheng, Mingchen Ma, Yihang Zou, Ningyuan Yang, Wei Cheng, \\
Hai ``Helen'' Li, Yiran Chen, and Tingjun Chen}
\affiliation{%
  \institution{\vspace{1.0mm}Department of Electrical and Computer Engineering, Duke University}
  \city{}
  \state{}
  \country{}
}

\renewcommand{\shortauthors}{Li et al.}

\begin{abstract}
Large language models (LLMs) and agentic systems have shown promise for automated software development, but applying them to hardware-in-the-loop (HIL) embedded and Internet-of-Things (IoT) systems remains challenging due to the tight coupling between software logic and physical hardware behavior.
Code that compiles successfully may still fail when deployed on real devices because of timing constraints, peripheral initialization requirements, or hardware-specific behaviors.
To address this challenge, we introduce a skills-based agentic framework for HIL embedded development together with IoT-SkillsBench, a benchmark designed to systematically evaluate AI agents in real embedded programming environments.
IoT-SkillsBench spans three representative embedded platforms, 23 peripherals, and 42 tasks across three difficulty levels, where each task is evaluated under three agent configurations (no-skills, LLM-generated skills, and human-expert skills) and validated through real hardware execution.
Across 378 hardware-validated experiments, we show that concise human-expert skills with structured expert knowledge enable near-perfect success rates across platforms.
\end{abstract}

\maketitle

%%%%%
%%%%%
\section{Introduction}
\label{sec:intro}

Recent advances in large language models (LLMs) and agentic systems have shown promise in assisting software development by enabling automated code generation~\cite{zhuo2024bigcodebench,jain2024livecodebench,song2026nemo}, debugging~\cite{jimenez2023swe}, and reasoning over complex documentation~\cite{yang2024swe}.
This capability has motivated growing interest in applying LLM-based agents to hardware-in-the-loop (HIL) embedded systems and Internet-of-Things (IoT) application development~\cite{shen2025gpiot,shen2025autoiot,gong2025programming,xu2025embedagent,yang2025iot,liu2025tasksense}.
However, unlike purely software-based agentic systems operating in cloud or local environments, designing agentic frameworks for embedded and IoT systems interacting with real physical hardware introduces fundamentally different and significantly more complex challenges.
% These characteristics suggest that naive prompt-based code generation, which treats embedded system development as a purely textual synthesis task, is fundamentally insufficient.

% \myparatight{Device heterogeneity and tightly coupled hardware-software stacks.}
First, \emph{the embedded ecosystem is highly fragmented}. 
Microcontroller units (MCUs) rely on diverse vendor-specific toolchains and real-time operating system (RTOS) environments (e.g., ESP-IDF, Zephyr, AVR-Libc), while peripheral interactions span diverse communication protocols (e.g., I2C, SPI, UART, wireless stacks).
Beyond heterogeneity, interfacing with physical hardware requires strict adherence to low-level constraints, including precise timing, initialization ordering, interrupt handling, and pin multiplexing.
Moreover, embedded platforms exhibit tight coupling between hardware and software, and agents must navigate across hardware-specific configurations and software behaviors, where even minor configuration mismatches can result in silent system failures.
Existing approaches mitigate this challenge by augmenting the agent's memory~\cite{thakur2025freshstack} with library documentation~\cite{yang2024embedgenius} or by resolving conflicts between headers/APIs and their usage through state flow graph analysis~\cite{gong2025programming}.
However, they substantially expand the required context window, thereby incurring significant token usage~\cite{chen2023frugalgpt}.

Second, \emph{for embedded systems, successful compilation and flashing to a device do not guarantee correct runtime behavior.}
Code that builds without errors may still exhibit unexpected behavior due to timing violations, peripheral misconfigurations, or hardware-specific edge cases. 
This gap arises because hardware behavior cannot be fully or deterministically validated through software-only verification techniques.
Although hardware emulation frameworks such as QEMU~\cite{qemu_website} and Wokwi~\cite{Wokwi} exist, they suffer from limited peripheral fidelity, incomplete timing models, and poor support for real-world sensor-actuator interactions. 
As a result, many hardware faults cannot be captured in purely digital emulation environments. 
Existing works that rely on compiler-in-the-loop~\cite{xu2025embedagent} or flashing-in-the-loop~\cite{yang2024embedgenius} paradigms remain limited, as they fail to incorporate behavioral errors that emerge during physical execution.

\begin{figure}[!t]
    \centering
    \includegraphics[width=0.98\linewidth]{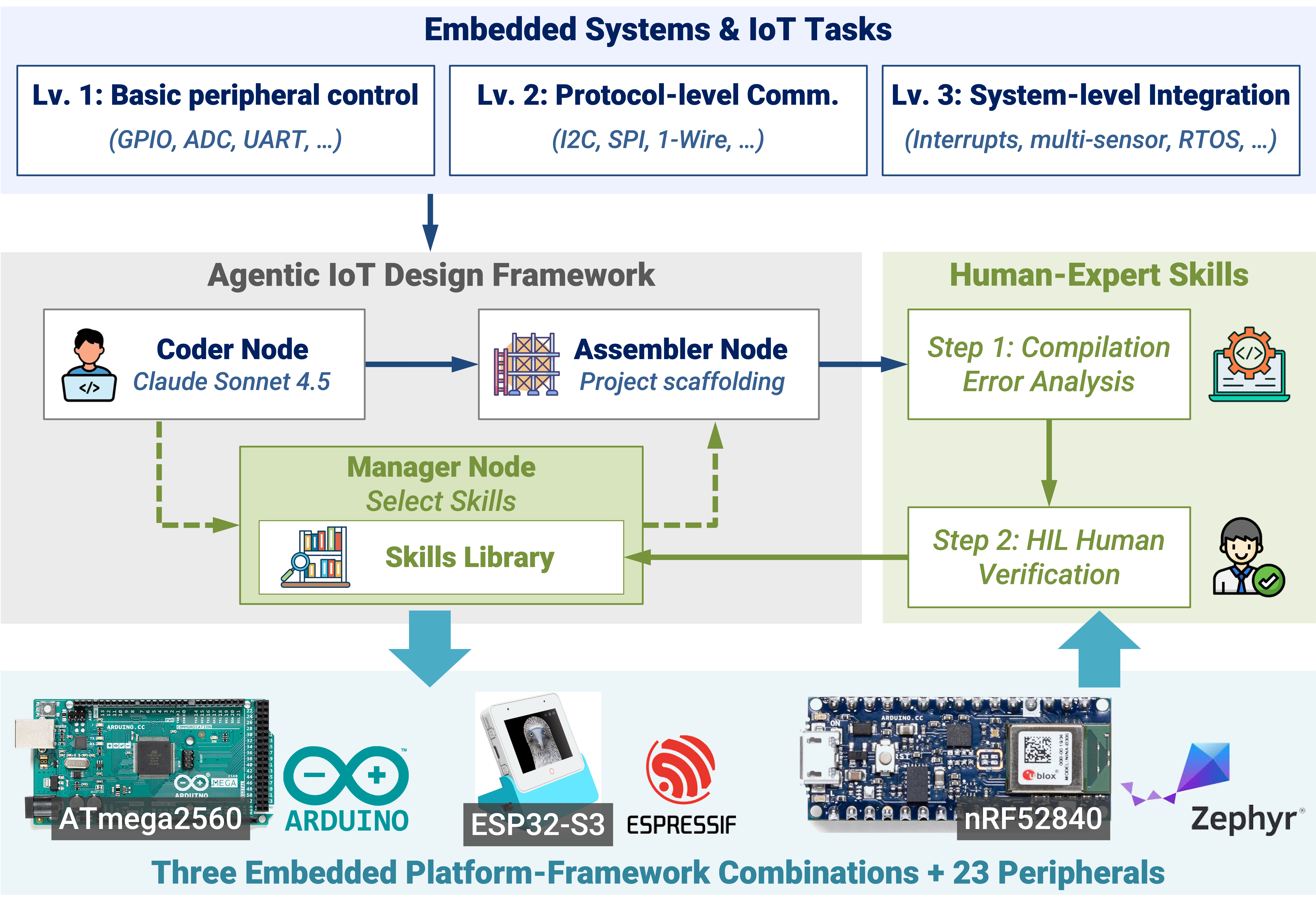}
    \vspace{-2.0mm}
    \caption{Overview of the skills-based agentic framework for embedded and IoT systems development and the {\name} benchmark.}
    \label{fig:system-overview}
    \vspace{-3.0mm}
\end{figure}

To address these challenges, we propose a \emph{skills-based agentic framework for embedded systems and IoT application development} (Fig.~\ref{fig:system-overview}).
Rather than injecting large volumes of SDK documentation or datasheets into the prompt, each skill distills the essential programming patterns, initialization constraints, and known failure modes for a specific peripheral, MCU, or framework combination into a compact, human-readable document.
This allows the agent to reason over a constrained and validated set of knowledge rather than raw documentation or APIs, significantly reducing token overhead while improving reliability.
Critically, the skills-based design~\cite{li2026skillsbench} scales naturally as the embedded ecosystem grows: adding support for a new platform, peripheral, or framework requires only authoring the corresponding skills, while the agent pipeline and evaluation harness remain unmodified.

To evaluate the role of structured hardware knowledge in agentic embedded development, we introduce \emph{\name}, a HIL benchmark covering multiple MCU platforms, development frameworks, and task complexities.
{\name} spans three representative platform-framework combinations (ATmega2560+Arduino, ESP32-S3+ESP-IDF, nRF52840+Zephyr) and contains 42 tasks across three difficulty levels: basic peripheral control, protocol-level communication (e.g., I2C, SPI), and system-level integration involving interrupts and multi-device coordination.
Each task is evaluated under three agent configurations: no-skills, LLM-generated skills, and human-expert skills, enabling controlled comparison between raw LLM knowledge, automatically synthesized knowledge, and carefully curated domain expertise.
In total, 378 evaluation instances across platforms, tasks, and skills conditions are validated through real hardware execution and behavioral tests.

Experiments on {\name} reveal three key insights.
First, modern LLMs already performed well on simple embedded tasks for well-documented platforms such as Arduino, but performance degrades rapidly as tasks require protocol-level reasoning or system integration. 
Second, LLM-generated skills provide inconsistent benefits and can even degrade performance, often reinforcing incorrect platform-specific assumptions while significantly increasing token usage.
Third, concise human-expert skills dramatically improved reliability, achieving near-perfect success rates across platforms and task levels while maintaining moderate token overhead.
These results highlight that the effectiveness of agentic embedded development depends not merely on providing skills, but on the quality and grounding of those skills in real hardware behavior, suggesting that structured, expert-curated hardware knowledge remains essential for reliable HIL software generation.

{\name} is open-source and available at~\cite{github_repo}.

%%%%%
%%%%%
\section{{\name}: Platforms, Tasks, and Skills}
\label{platforms}
\myparatight{Embedded system platforms and peripherals.}
We consider three platform-framework combinations, where each platform refers to a specific MCU development board paired with its corresponding embedded framework:
\begin{itemize}[leftmargin=*, topsep=2pt, itemsep=1pt]
    \item
    \textbf{ATmega2560} (Arduino Mega 2560 Rev3) with the Arduino framework (Arduino CLI v1.4.1, Arduino Core arduino:avr v1.8.7);
    \item
    \textbf{ESP32-S3} (ESP32-S3-BOX-3) with ESP-IDF (v5.1.2);
    \item
    \textbf{nRF52840} (Arduino Nano 33 BLE Rev2) with Zephyr via nRF Connect SDK (v2.7.0).
\end{itemize}
These platforms were selected to cover different MCU architectures (AVR, Xtensa, ARM Cortex-M4), HAL paradigms (Arduino-style abstraction, bare-metal register access, RTOS-based drivers), and development toolchain ecosystems.
Together, they represent a broad cross-section of the modern embedded systems development landscape.
In addition to Arduino that has been considered in recent works (e.g.,~\cite{yang2024embedgenius}), our framework also supports ESP-IDF and Zephyr, which are two widely used industrial embedded development frameworks that power many commercial IoT and edge devices, thereby enabling evaluation in settings that more closely reflect real-world embedded engineering workflows.

We consider 23 peripherals as shown in Appendix~\ref{app:sensors_list}, covering a wide range of hardware interfaces from simple actuators and indicators (e.g., LEDs, buzzers, relays, buttons) to sensors spanning analog output, digital protocols, and complex communication buses including I2C and SPI.
This peripheral diversity ensures that benchmark tasks exercise the full depth of the hardware-software interface rather than a narrow subset of peripheral types.

\myparatight{Embedded and IoT Development Tasks.}
We carefully construct a benchmark consisting of 42 tasks across three difficulty levels.
\emph{Level~1} tasks cover basic peripheral control, including GPIO toggling, ADC sampling, and UART communication.
\emph{Level~2} tasks require protocol-level reasoning over I2C, SPI, and 1-Wire buses, including correct initialization sequences and timing constraints.
\emph{Level~3} tasks demand system-level integration, such as interrupt-driven workflows, multi-sensor/actuator integrations, and (complex) task scheduling.
Specifically, each task is defined by:
(\emph{i}) a natural language-based task description, optionally specifying the peripheral or sensor to be used and the logic to be implemented (e.g., ``debouncing for a button input''); and
(\emph{ii}) a target hardware platform and framework (e.g., ``nRF52840 + Zephyr'') with the required pin assignments for all connected peripherals.
Note that pin referencing conventions may differ across platforms (e.g., ESP-IDF uses numeric GPIO numbers while Zephyr uses device tree aliases), and these platform-specific pin descriptors are provided to ensure the agent can generate compilable code directly without needing to resolve hardware mappings independently.
Each task was manually validated on physical hardware with human observations prior to inclusion, and reference implementations were verified to produce correct HIL test outcomes.
All tasks are guaranteed to have a valid reference implementation on their corresponding target platforms.

%%%%%
%%%%%
\myparatight{Skills specifications.}
We construct two skills sets for the agent.
The first set is generated by prompting a modern LLM (Claude Sonnet 4.6) with each task and its corresponding platform-framework combinations.
The LLM is asked to summarize the relevant embedded programming knowledge purely from its parametric knowledge, without access to datasheets, error logs, or external documentation.
The second set is carefully constructed by embedded systems experts with hands-on experience on the target platforms and peripherals. 
The goal is to produce skills that span diverse hardware platforms and peripherals while remaining \emph{compact, targeted, and directly actionable for the agent}.
During the skill authoring process, human experts are provided with the LLM-generated embedded code for each task (produced without skills), along with the corresponding compiler error logs and observed runtime behaviors.
This provides a grounded view of the typical failure modes of LLM-generated embedded code.
Moreover, experts are instructed to write skills concisely, favoring descriptive prose over code snippets, and to keep each skill focused on a single peripheral or framework-specific combination.
Both skill sets follow a unified format inspired by the Agent Skills specification~\cite{agent_skills_github}, where each skill file consists of a YAML header describing the skill's metadata (e.g., name, target platform, or peripherals) followed by a Markdown body containing the skill content itself.
This structured format allows the agent to efficiently scan skill headers during retrieval without loading the full skill body, therefore keeping token overhead low.

\myparatight{Summary.}
Overall, {\name} combines three platform-framework pairs and 42 tasks, and three skills configurations (no-skills, LLM-generated skills, and human-expert skills).
This yields a total number of $42 \times 3 = 126$ evaluation instances per platform, and $126 \times 3 = 378$ instances across the entire benchmark.
This design allows direct comparison between no-knowledge, LLM-synthesized knowledge, and human-expert knowledge, therefore isolating the contribution of skills quality.
This makes {\name} useful not only as a benchmark for embedded agent performance, but also as a testbed for studying how procedural knowledge affects agent behavior in hardware-constrained settings.

%%%%%
%%%%%
\section{Agent Setup and Evaluation Protocol}

We implement the agent using LangGraph~\cite{langgraph} with a three-node architecture consisting of a \emph{manager node}, a \emph{coder node}, and an \emph{assembler node}.
We intentionally adopt this minimal structure to isolate the effect of skills from other complexities of agentic system design.
For example, many modern agentic frameworks incorporate multi-step planning, tool use, iterative debugging, or reflective loops, which can confound the evaluation of whether improvements arise from skills themselves or from additional reasoning and orchestration mechanisms.
By constraining the agent to a single-pass pipeline with only one optional planning stage, our evaluation focuses specifically on the contribution of skills for HIL systems.

%% table begins
\begin{table}[!t]
\centering
\caption{Detailed {CF/BF/BC}@1 outcomes across platforms, skills configurations, and task difficulty levels.
Each cell reports the number of CF/BF/BC tasks.}
\label{tab:main-results-pass1}
\vspace{-2.0mm}
\small
\resizebox{\linewidth}{!}{
\begin{tabular}{cccccc}
\hline\noalign{\vspace{2pt}}
\textbf{Skills} & \textbf{Platform} & \textbf{L1 (12)} & \textbf{L2 (16)} & \textbf{L3 (14)} & \textbf{Total (42)} \\
\hline\noalign{\vspace{2pt}}
\multirow{3}{*}[-2pt]{No-Skills}
  & Arduino & 0/0/12 & 1/1/14 & 2/1/11 & 3/2/37 \\[1pt]
  & ESP-IDF & 0/0/12 & 4/3/9 & 1/8/5 & 5/11/26 \\[1pt]
  & Zephyr  & 0/2/10 & 1/5/10 & 2/8/4  & 3/15/24 \\
\hline\noalign{\vspace{2pt}}
\multirow{3}{*}[-2pt]{LLM}
  & Arduino & 0/0/12 & 1/0/15 & 1/1/12 & 2/1/39 \\[1pt]
  & ESP-IDF & 0/0/12 & 1/4/11 & 0/11/3 & 1/15/26 \\[1pt]
  & Zephyr  & 0/2/10 & 4/8/4  & 0/9/5 & 4/19/19 \\
\hline\noalign{\vspace{2pt}}
\multirow{3}{*}[-2pt]{\begin{tabular}{@{}c@{}}Human-\\Expert\end{tabular}}
  & Arduino & 0/0/12 & 0/0/16 & 0/1/13 & 0/1/41 \\[1pt]
  & ESP-IDF & 0/0/12 & 0/1/15 & 0/1/13 & 0/2/40 \\[1pt]
  & Zephyr  & 0/0/12 & 0/2/14 & 0/1/13 & 0/3/39 \\
\hline
\end{tabular}
}
\vspace{-4.0mm}
\end{table}
%% table ends

When no skills are provided, the manager node is skipped, and the task prompt is directly passed to the coder node.
When skills are enabled, the manager node receives a system prompt instructing it to act as a project planner, along with the list of all available skill headers and task requirements.
Based on this, it selects the relevant skills and passes them to the coder node along with the task prompt.
The coder node receives a system prompt instructing it to act as an expert embedded engineer, the selected skill content as applicable standards, and the task requirements as the user message.
It then generates the main firmware code file depending on the target framework (e.g., \myCodeShort{*.c} for ESP-IDF and Zephyr or \myCodeShort{*.ino} for Arduino).

Finally, the assembler node formats the raw LLM output into a ready-to-compile project, extracting clean code and generating all necessary platform-specific scaffolding. For ESP-IDF and Zephyr, this includes top-level and component-level \myCodeShort{CMakeLists.txt} files to define build targets. For Zephyr, it additionally generates a \myCodeShort{prj.conf} file for configuration options (e.g., enabling I2C/SPI), and a device tree overlay (\myCodeShort{*.overlay}) to map peripherals to device tree nodes. This separation ensures the coder focuses purely on firmware logic while the assembler handles all platform-specific project structure.

%%%%%
%%%%%
\myparatight{Evaluation protocol.}
We use Claude Sonnet~4.5 as the backbone model for all experiments across the benchmark, and the agent executes each task-skills-platform combination five times.
To eliminate any variability introduced by toolchain differences, all generated project folders are compiled manually using the exact same toolchain commands and identical compiler configurations across all tasks for the targeted platforms (see \S\ref{platforms} for the exact toolchain and framework versions used).
The compiled firmware is then flashed and tested on the physical target hardware, and the results are recorded and validated by a human evaluator.
This process ensures that the evaluation reflects the true behavior of the deployed firmware on real physical hardware rather than artifacts of the build system or simulation environment.
\emph{Due to the HIL nature of the benchmark, the full evaluation required approximately 100 hours of human-in-the-loop validation across all tasks.}

%% table begins
\begin{table}[!t]
\centering
\caption{Detailed {CF/BF/BC}@5 outcomes across platforms, skills configurations, and task difficulty levels.
Each cell reports the number of CF/BF/BC tasks.}
\label{tab:main-results-pass5}
\vspace{-2.0mm}
\small
\resizebox{\linewidth}{!}{
\begin{tabular}{cccccc}
\hline\noalign{\vspace{2pt}}
\textbf{Skills} & \textbf{Platform} & \textbf{L1 (12)} & \textbf{L2 (16)} & \textbf{L3 (14)} & \textbf{Total (42)} \\
\hline\noalign{\vspace{2pt}}
\multirow{3}{*}[-2pt]{No-Skills}
  & Arduino & 0/0/12 & 0/0/16 & 0/0/14 & 0/0/42 \\[1pt]
  & ESP-IDF & 0/0/12 & 2/2/12 & 1/6/7  & 3/8/31 \\[1pt]
  & Zephyr  & 0/1/11 & 0/5/11 & 0/8/6  & 0/14/28 \\
\hline\noalign{\vspace{2pt}}
\multirow{3}{*}[-2pt]{LLM}
  & Arduino & 0/0/12 & 0/0/16 & 1/0/13 & 1/0/41 \\[1pt]
  & ESP-IDF & 0/0/12 & 1/4/11 & 0/10/4 & 1/14/27 \\[1pt]
  & Zephyr  & 0/2/10 & 1/6/9  & 0/6/8  & 1/14/27 \\
\hline\noalign{\vspace{2pt}}
\multirow{3}{*}[-2pt]{\begin{tabular}{@{}c@{}}Human-\\Expert\end{tabular}}
  & Arduino & 0/0/12 & 0/0/16 & 0/0/14 & 0/0/42 \\[1pt]
  & ESP-IDF & 0/0/12 & 0/0/16 & 0/1/13 & 0/1/41 \\[1pt]
  & Zephyr  & 0/0/12 & 0/1/15 & 0/0/14 & 0/1/41 \\
\hline
\end{tabular}
}
\vspace{-4.0mm}
\end{table}
%% table ends

\myparatight{Performance metrics.}
Each task attempt will lead to one of three following outcomes:
\begin{itemize}[leftmargin=*, topsep=2pt, itemsep=1pt]
    \item
    \textbf{Compile Failure (CF)}: The generated firmware fails to compile or cannot be flashed to the target device;
    \item
    \textbf{Behavior Failure (BF)}: The firmware compiles and flashes successfully but fails the hardware behavioral test, including cases where the device hangs or triggers a watchdog reset;
    \item
    \textbf{Behavior Correct (BC)}: The firmware compiles, flashes, and produces the correct observable hardware behavior as verified by the HIL test harness.
\end{itemize}
In addition to the breakdown of \textbf{\{CF, BF, BC\}} for each task attempt, we consider two aggregate metrics:
\textbf{pass@1} measures the fraction of tasks that achieve BC on the very first attempt; and
\textbf{pass@5} (equivalently, BC@5) measures the fraction of tasks where at least one of the five attempts achieves BC.
Together, they capture both the practical efficiency and the achievable ceiling performance of each task-skills-platform combination.
We also record \textbf{token usage} for each LLM call by extracting input and output token counts separately from the model's response metadata.
This allows us to report the total token cost per task broken down by node (manager and coder), and to directly compare the token overhead introduced by skill-augmented conditions against the no-skills baseline.

%%%%%
%%%%%
\section{Experimental Results}

\begin{figure}[!t]
    \centering
    \includegraphics[width=0.95\linewidth]{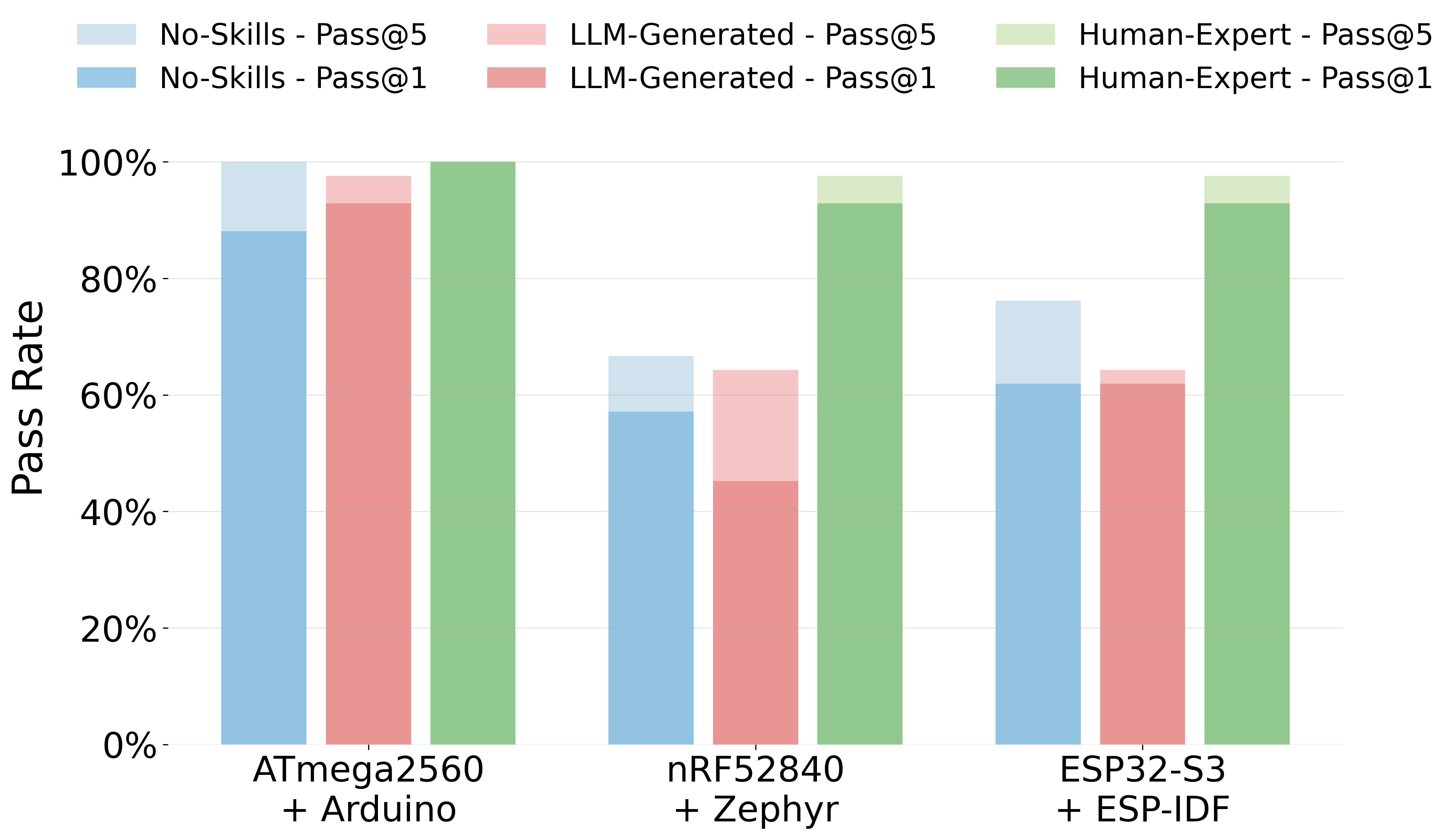}
    \vspace{-3.0mm}
    \caption{Pass rates comparison (Pass@1 and Pass@5) across platforms, skills configurations, and task difficulty levels.}
    \label{fig:pass-rates}
    \vspace{-3.0mm}
\end{figure}

\begin{figure}[!t]
    \centering
    \includegraphics[width=0.95\linewidth]{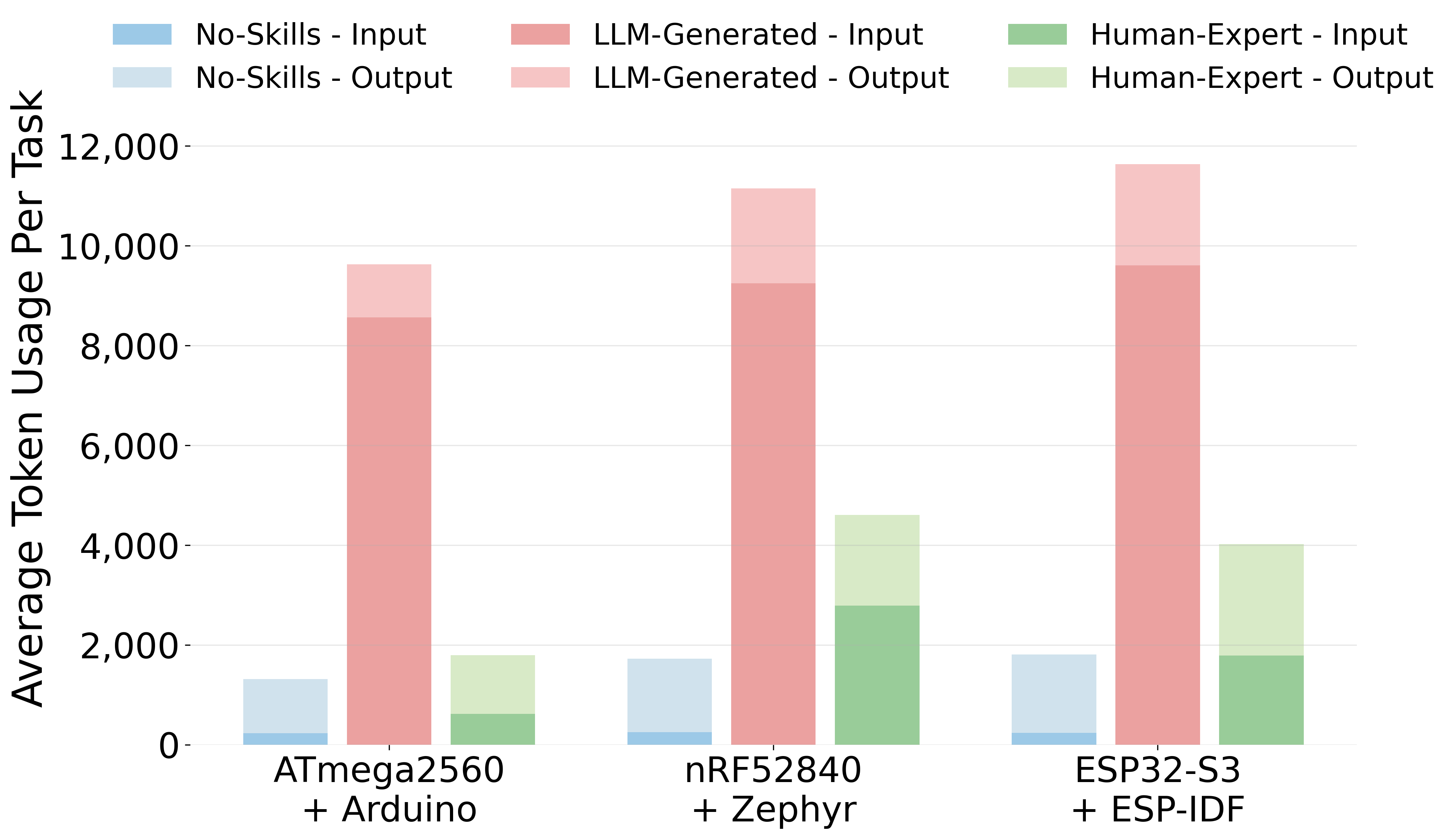}
    \vspace{-3.0mm}
    \caption{Average per-task token usage across platforms, skills configurations, and task difficulty levels.}
    \label{fig:token-usage}
    \vspace{-3.0mm}
\end{figure}

Tables~\ref{tab:main-results-pass1} and~\ref{tab:main-results-pass5}, Figs.~\ref{fig:pass-rates} and~\ref{fig:token-usage} report the main performance and token usage, respectively.
While the no-skills baseline resolves simple Level~1 tasks via pretraining knowledge, adding LLM-generated skills provides inconsistent benefits, suggesting that automatically synthesized skills with minimal guidelines may introduce noise or reinforce incorrect platform-specific assumptions.
In contrast, expert-authored skills consistently achieve near-perfect performance across all platforms and task difficulty levels, demonstrating that concise, expert-curated skills are both more effective and token-efficient than their LLM-generated counterparts.

\myparatight{No-skills baseline.}
Without any skills, the agent achieves perfect performance on Level~1 tasks for Arduino and ESP-IDF (12/12 each), but struggles on Zephyr (11/12).
This difference is likely due to the relative scarcity of Zephyr-specific material in LLM pretraining data compared with Arduino and ESP-IDF, both of which have extensive online resources, tutorials, and open-source examples.
Performance degrades substantially as task difficulty level increases across all platforms.
For Level~3 tasks, the agent resolves only 7/14 tasks on ESP-IDF and 6/14 tasks on Zephyr.
Overall, Zephyr emerges as the most challenging platform, with only 28/42 tasks resolved, compared with 42/42 for Arduino and 31/42 for ESP-IDF.
Token consumption in this baseline is minimal, averaging approximately 300 input tokens and 1,200 output tokens per task.

\myparatight{LLM-generated skills.}
Using LLM-generated skills yields mixed results across platforms.
For Arduino, performance shows only a minor degradation (41/42 vs. 42/42), suggesting that for a well-documented platform like Arduino, LLM-generated skills provide little additional value over the model's existing knowledge.
In contrast, ESP-IDF experiences a more noticeable performance drop from 31/42 to 27/42 resolved tasks, with Level~3 performance falling from 8/14 to 4/14.
This indicates that LLM-synthesized skills can sometimes reinforce incorrect assumptions about complex ESP-IDF-specific behavior.
Zephyr also exhibits a slight overall performance degradation (27/42 vs. 28/42), though Level~3 task performance improves modestly from 6/14 to 8/14.
This suggests that LLM-generated skills may provide some benefit for the most challenging Zephyr tasks.
Overall, LLM-generated skills do not consistently outperform the no-skills baseline and in several cases lead to degraded performance.
The agent with these skills also incurs the highest token cost, averaging approximately {8,500--9,500} input tokens and {1,500--2,000} output tokens per task.
Notably, despite instructing the coder node to output code only, output token counts remain elevated.
This is a consequence of the model reiterating or reasoning aloud about the skill guidelines before producing the final code.

\myparatight{Human-expert skills.}
Expert-crafted skills achieve the best overall performance across all platforms and task difficulty levels, with Arduino reaching a perfect 42/42 and both ESP-IDF and Zephyr achieving 41/42 resolved tasks.
The only unresolved ESP-IDF task involves a 5V RTC module causing instability on the 3.3V ESP32-S3, a hardware-level voltage incompatibility that cannot be resolved through firmware alone.
The only unresolved Zephyr task involves a rotary encoder where clockwise/counterclockwise direction behavior is not standardized across encoder models, making it impossible to specify a universally correct solution in the prompt.
These two failures represent fundamental hardware ambiguities rather than agent limitations.
Token consumption with human-expert skills remains moderate, averaging approximately {650--2,900} input tokens and {1,700--4,600} output tokens per task, striking a better balance between performance and efficiency compared to the LLM-generated skills setting.

%%%%%
%%%%%

%%%%%
%%%%%

%%%%%
%%%%%
\section{Conclusions}
In this work, we presented {\name}, a hardware-validated benchmark for evaluating agentic systems for embedded and IoT systems development, together with a skills-based agent framework that integrates structured procedural knowledge into the code generation pipeline.
Extensive HIL evaluations across multiple MCU platforms and task difficulty levels show that raw LLM capabilities alone are insufficient for reliable embedded development, while compact human-expert skills significantly improved reliability, achieving near-perfect task success while maintaining moderate token cost.
These results highlight that grounded domain knowledge---not merely larger models---is essential for agentic programming in hardware-constrained environments.
We hope {\name} will serve as a foundation for future research on AI-assisted embedded development, HIL agent evaluation, and scalable knowledge representations for physical-system programming.

\section{Acknowledgments}
The work was supported in part by NSF grant CNS-2112562.

\newpage
\bibliographystyle{ACM-Reference-Format}
\bibliography{ref}

%%%%%
%%%%%
\newpage

\appendix
\section{List of Peripherals}
\label{app:sensors_list}

Table~\ref{tab:peripherals} lists the peripherals supported by {\name}, all validated through HIL testing.

\begin{table}[h]
\centering
\caption{List of peripherals supported by {\name}.}
\label{tab:peripherals}
\vspace{-2.0mm}
\small
\setlength{\tabcolsep}{4pt}
\renewcommand{\arraystretch}{1.2}
\resizebox{\linewidth}{!}{
\begin{tabular}{clll}
\toprule
\textbf{\#} & \textbf{Peripheral} & \textbf{Interface} & \textbf{Category} \\
\midrule
1  & LED                           & GPIO (Digital Out)  & Actuator \\
2  & Push Button                   & GPIO (Digital In)   & Input    \\
3  & Active Buzzer                 & GPIO (Digital Out)  & Actuator \\
4  & Passive Buzzer                & PWM                 & Actuator \\
5  & Relay Module                  & GPIO (Digital Out)  & Actuator \\
6  & Laser Emitter Module          & GPIO (Digital Out)  & Actuator \\
7  & Rotary Encoder                & GPIO (Digital In)   & Input    \\
8  & 16-Key Keypad (4$\times$4)    & GPIO (Digital In)   & Input    \\
9  & Tilt Switch (KY-020)          & GPIO (Digital In)   & Input    \\
10 & Analog Joystick               & ADC                 & Input    \\
11 & Photoresistor (KY-018)        & ADC                 & Sensor   \\
12 & TMP36 Temperature Sensor      & ADC                 & Sensor   \\
13 & Analog Water Level Sensor     & ADC                 & Sensor   \\
14 & PIR Motion Sensor (HC-SR501)  & GPIO (Digital In)   & Sensor   \\
15 & Ultrasonic Sensor (HC-SR04)   & GPIO (Trigger/Echo) & Sensor   \\
16 & Digital Sound Sensor          & GPIO (Digital In)   & Sensor   \\
17 & Digital Shock Sensor          & GPIO (Digital In)   & Sensor   \\
18 & DHT11 (Temp \& Humidity)      & 1-Wire              & Sensor   \\
19 & DS18B20 (Temperature)         & 1-Wire              & Sensor   \\
20 & LCD1602 Display (HD44780)     & GPIO (Parallel)  & Output   \\
21 & DS1307 RTC Module             & I2C                 & Sensor   \\
22 & MPU6050 / GY-521              & I2C                 & Sensor   \\
23 & BME280 (Temp, Humidity, Pres.)& I2C / SPI           & Sensor   \\
\bottomrule
\end{tabular}
}
\end{table}
%%

%%%%%
%%%%%
\section{Prompt for LLM-based Skills Generation}

The following prompt is used to generate the LLM-generated skills:

\begin{tcolorbox}[
    enhanced,
    colframe=darkgray,
    colback=gray!5!white,
    coltitle=white,
    fonttitle=\bfseries,
    fontupper=\ttfamily\small\raggedright, % <--- Added \raggedright here!
    title={Prompt: LLM-based Skills Generation},
    arc=1mm,
    boxrule=1.0pt,
    left=4pt, right=4pt, top=4pt, bottom=4pt
]
Read tasks in tasks/level<X>.txt, please follow these steps:\\[1ex]
1. Analyze the task requirements and identify what domain knowledge, hardware platforms, or connectivity protocols, etc, are needed.\\[1ex]
2. Consider implementation based on three MCU + development framework combinations: 1) ESP32 + ESP-IDF, 2) ATMega2560 + Arduino, 3) nRF52840 + Zephyr.\\[1ex]
3. Write modular skill documents that would help solve these tasks. Each skill should: focus on a specific MCU, board, IoT tool, protocol, framework, or embedded technique; provide code examples and usage patterns; be reusable for similar tasks.\\[1ex]
4. Save each skill as a markdown file in the skills-llm-generated/ directory with a descriptive name.
\end{tcolorbox}

%%%%%
%%%%%
\section{Example Tasks Across Difficulty Levels}

Below, we provide example tasks in {\name} across three task difficulty levels.

\subsection{Level 1: Basic Peripheral Control}

\vspace{0.5ex}
\begin{tcolorbox}[
    enhanced,
    colframe=MidnightBlue,
    colback=MidnightBlue!5!white,
    coltitle=white,
    fonttitle=\bfseries,
    fontupper=\ttfamily\small,
    title={Task: Button Status with Debouncing},
    arc=1mm,
    boxrule=1.0pt,
    left=4pt, right=4pt, top=4pt, bottom=4pt
]
Read the state of a pull-down button and implement software debouncing to avoid multiple triggers. Print "Button Pressed!" to the serial console when the button is pressed.
\end{tcolorbox}

\begin{tcolorbox}[
    enhanced,
    colframe=MidnightBlue,
    colback=MidnightBlue!5!white,
    coltitle=white,
    fonttitle=\bfseries,
    fontupper=\ttfamily\small,
    title={Task: TMP36},
    arc=1mm,
    boxrule=1.0pt,
    left=4pt, right=4pt, top=4pt, bottom=4pt
]
Read the analog output of the TMP36 temperature sensor, where the output voltage is linearly proportional to the temperature in degrees Celsius, and print the temperature value to the serial console.
\end{tcolorbox}

\begin{tcolorbox}[
    enhanced,
    colframe=MidnightBlue,
    colback=MidnightBlue!5!white,
    coltitle=white,
    fonttitle=\bfseries,
    fontupper=\ttfamily\small,
    title={Task: SOS Morse Code},
    arc=1mm,
    boxrule=1.0pt,
    left=4pt, right=4pt, top=4pt, bottom=4pt
]
Blink the LED to spell out “SOS” in Morse code. 
\end{tcolorbox}

\subsection{Level 2: Protocol-Level Communication}

\vspace{0.5ex}
\begin{tcolorbox}[
    enhanced,
    colframe=MidnightBlue,
    colback=MidnightBlue!5!white,
    coltitle=white,
    fonttitle=\bfseries,
    fontupper=\ttfamily\small,
    title={Task: MPU6050 Data Reading (I2C)},
   arc=1mm,
    boxrule=1.0pt,
    left=4pt, right=4pt, top=4pt, bottom=4pt
]
Read raw accelerometer and gyroscope data from the MPU6050 via I2C and print to the serial console. 
\end{tcolorbox}

\begin{tcolorbox}[
    enhanced,
    colframe=MidnightBlue,
    colback=MidnightBlue!5!white,
    coltitle=white,
    fonttitle=\bfseries,
    fontupper=\ttfamily\small,
    title={Task: BME280 Data Reading (SPI)},
    arc=1mm,
    boxrule=1.0pt,
    left=4pt, right=4pt, top=4pt, bottom=4pt
]
Read the humidity and temperature from a BME280 sensor using the SPI bus, and print the humidity and temperature values to the serial console. 
\end{tcolorbox}

\begin{tcolorbox}[
    enhanced,
    colframe=MidnightBlue,
    colback=MidnightBlue!5!white,
    coltitle=white,
    fonttitle=\bfseries,
    fontupper=\ttfamily\small,
    title={Task: LCD1602 Display "Hello World"},
    arc=1mm,
    boxrule=1.0pt,
    left=4pt, right=4pt, top=4pt, bottom=4pt
]
Set up the LCD1602 display, and display "Hello World" at the center of the screen. 
\end{tcolorbox}

\subsection{Level 3: System-Level Integration}

\vspace{0.5ex}
\begin{tcolorbox}[
    enhanced,
    colframe=MidnightBlue,
    colback=MidnightBlue!5!white,
    coltitle=white,
    fonttitle=\bfseries,
    fontupper=\ttfamily\small,
    title={Task: Safe Box with Display},
    arc=1mm,
    boxrule=1.0pt,
    left=4pt, right=4pt, top=4pt, bottom=4pt
]
Write the program that will read the password input from the 16 key keypad, there are in total 8 lines connected to each 4 rows and 4 cols. In particular, the password will be set to "1234". The program will read the key input from the 16 key keypad, and if the input matches the password, the program will connect the relay to unlock the safebox.
\end{tcolorbox}

\begin{tcolorbox}[
    enhanced,
    colframe=MidnightBlue,
    colback=MidnightBlue!5!white,
    coltitle=white,
    fonttitle=\bfseries,
    fontupper=\ttfamily\small,
    title={Task: Automatic Brightness Control for LCD1602 Display},
    arc=1mm,
    boxrule=1.0pt,
    left=4pt, right=4pt, top=4pt, bottom=4pt
]
Use the ambient light intensity (the KY-018 photoresistor) to automatically adjust the brightness of the LCD1602 backlight. Read the analog value from the KY-018 photoresistor, map it to a suitable PWM duty cycle, and control the backlight brightness accordingly.
\end{tcolorbox}

\begin{tcolorbox}[
    enhanced,
    colframe=MidnightBlue,
    colback=MidnightBlue!5!white,
    coltitle=white,
    fonttitle=\bfseries,
    fontupper=\ttfamily\small,
    title={Task: Variable Frequency LED with Buzzer},
    arc=1mm,
    boxrule=1.0pt,
    left=4pt, right=4pt, top=4pt, bottom=4pt
]

Build the program that integrates the button press with both the buzzer and the timer-based LED control. After the system reset, when you press the button:

- For the 1st time, a timer is triggered that toggles the external LED at 1 Hz;

- For the 2nd time, a timer is triggered that toggles the external LED at 2 Hz;

- For the 3rd time, a timer is triggered that toggles the external LED at 4 Hz;

- For the 4th time, the timer is stopped and the external LED will not blink;

The process repeats and the toggling frequency of the LED will undergo the sequence of 1 Hz, 2 Hz, 4 Hz, N/A, 1 Hz, 2 Hz, 4 Hz, N/A, … as you press the button. In addition, every time the button is pressed, the buzzer will go off, indicating that the button has been pressed. The button and buzzer must be connected to separate GPIO pins, and the buzzer operation will be triggered by its own connected GPIO pin.

\end{tcolorbox}

\section{Human-Expert Skills Examples}

Below, we provide example human-expert skills in {\name} for GPIO best practices.

\begin{tcolorbox}[
    enhanced,
    colframe=teal,
    colback=teal!5!white,
    coltitle=white,
    fonttitle=\bfseries,
    fontupper=\ttfamily\small,
    title={Zephyr Skills: GPIO Best Practices},
    arc=1mm,
    boxrule=1.0pt,
    left=4pt, right=4pt, top=4pt, bottom=4pt
]
\begin{itemize}[leftmargin=*, nosep]
  \item Define pin polarity (\texttt{GPIO\_ACTIVE\_HIGH}/\texttt{GPIO\_ACTIVE\_LOW}) in the devicetree, not in code.
  \item Use \texttt{gpio\_pin\_set(dev, pin, 1)}, logical ON to turn ON; \texttt{gpio\_pin\_set(dev, pin, 0)}, logical OFF to turn OFF.
        Let Zephyr's GPIO driver handle the physical HIGH/LOW translation based on DT flags. Never assume physical voltage level.
  \item Use \texttt{GPIO\_INT\_EDGE\_TO\_ACTIVE}/\texttt{GPIO\_INT\_EDGE\_TO\_INACTIVE} instead of RISING/FALLING --- stays polarity-agnostic.
\end{itemize}
...
\end{tcolorbox}

\vspace{0.5ex}
\begin{tcolorbox}[
    enhanced,
    colframe=teal,
    colback=teal!5!white,
    coltitle=white,
    fonttitle=\bfseries,
    fontupper=\ttfamily\small,
    title={ESP-IDF Skills: GPIO Best Practices},
    arc=1mm,
    boxrule=1.0pt,
    left=4pt, right=4pt, top=4pt, bottom=4pt
]
\begin{itemize}[leftmargin=*, nosep, itemsep=4pt]
  % \item \textbf{Batch Configuration:} Avoid using consecutive \texttt{gpio\_set\_direction()} calls to initialize multiple pins. Always use \texttt{gpio\_config\_t} with a \texttt{pin\_bit\_mask} to configure pins simultaneously.

  \item \textbf{Prevent Floating States:} Explicitly configure internal resistors (\texttt{.pull\_up\_en}, \texttt{.pull\_down\_en}). If a pin must default to LOW/HIGH, use \texttt{gpio\_set\_level()} immediately after \texttt{gpio\_config()} to guarantee a deterministic state before peripheral power-on delays.

  % \item \textbf{Microsecond State Switching:} Dynamically switching between \texttt{GPIO\_MODE\_INPUT} and \texttt{GPIO\_MODE\_OUTPUT} during microsecond bit-banging is too slow and acquires locks. Instead, configure the pin once as \texttt{GPIO\_MODE\_INPUT\_OUTPUT\_OD} (Open-Drain). To send a LOW signal, output \texttt{0}. To release the bus for input or a HIGH signal, output \texttt{1} and read the state via \texttt{gpio\_get\_level()}.

  \item \textbf{Precise Timing \& Delays:} Use \texttt{ets\_delay\_us()} for microsecond-level precision; however, always include \texttt{\#include "rom/ets\_sys.h"} to prevent implicit declaration errors in modern IDF versions.

  \item \textbf{Interrupts for Transients:} For catching fast, transient state changes (e.g., \textless{}5ms pulses), avoid \texttt{while(1)} polling. Prefer Interrupt Service Routines via \texttt{gpio\_install\_isr\_service()} and configure edge-triggered interrupts (e.g., \texttt{GPIO\_INTR\_NEGEDGE}).
\end{itemize}
...
\end{tcolorbox}

\end{document}